\documentclass[floatfix,
reprint,
superscriptaddress,
nofootinbib,
nobibnotes,
 nofootinbib,amsmath,amssymb,
 aps,prl,
 twoside,
 lineno
]{revtex4-1}

\usepackage{graphicx}
\usepackage{dcolumn}
\usepackage{bm}
\usepackage{amsmath}
\usepackage{cases}
\usepackage{color}
\usepackage[columnwise]{lineno}

\newcommand{\thetahorz}{\theta_h}
\newcommand{\thetavert}{\theta_v}

\newcommand{\shape}{S}
\newcommand{\period}{T}
\newcommand{\rad}{R} 
\newcommand{\moment}{I} 
\newcommand{\pos}{\phi} 
\newcommand{\inner}{\alpha} 
\newcommand{\torque}{\tau} 
\newcommand{\omegagait}{\omega_g} 
\newcommand{\avepos}{\bar{\phi}}
\newcommand{\angmom}{L}

\newcommand{\mass}{m}
\newcommand{\pointi}{\hat{s}^{(i)}} 
\newcommand{\veli}{\partial_t{\hat{s}}^{(i)}} 
\newcommand{\rot}{\mathbf{R}} 
\newcommand{\angmomvec}{\mathbf{L}} 
\newcommand{\nmoment}[1]{\langle I^{(#1)} \rangle}
\newcommand{\vf}{\eta} 
\newcommand{\cf}{\tau_\textrm{C}} 
\newcommand{\spring}{\tau_g} 

\begin{document}

\preprint{APS/123-QED}

\title{Locomotion without force, and impulse via dissipation: Robotic swimming in curved space via geometric phase}

\author{Shengkai Li}
\thanks{These two authors contributed equally}
\affiliation{%
 School of Physics, Georgia Institute of Technology, Atlanta, GA 30332, USA}%
\author{Tianyu Wang}%
\thanks{These two authors contributed equally}
\affiliation{%
 School of Physics, Georgia Institute of Technology, Atlanta, GA 30332, USA}
\affiliation{Institute for Robotics and Intelligent Machines, College of Computing, Georgia Institute of Technology, Atlanta, GA, 30332, USA}%
 \author{Velin H. Kojouharov}%
\affiliation{%
 School of Mechanical Engineering, Georgia Institute of Technology, Atlanta, GA, 30332, USA}%
 \author{James McInerney}%
\affiliation{%
 Department of Physics, University of Michigan, Ann Arbor, MI 48109, USA}%
 \author{Yasemin O. Aydin}%
\affiliation{%
 Department of Electrical Engineering, University of Notre Dame, Notre Dame, IN 46556, USA}%
 \author{Enes Aydin}%
\affiliation{%
 Department of Electrical Engineering, University of Notre Dame, Notre Dame, IN 46556, USA}%
 \author{Daniel I. Goldman}%
\affiliation{%
 School of Physics, Georgia Institute of Technology, Atlanta, GA 30332, USA}%
\author{D. Zeb Rocklin}
\thanks{Correspondence and requests for materials should be addressed to D. Zeb Rocklin (zebrocklin@gatech.edu).}
\affiliation{%
 School of Physics, Georgia Institute of Technology, Atlanta, GA 30332, USA}%

\date{\today}

\begin{abstract}
Locomotion by shape changes (spermatozoon swimming, snake slithering, bird flapping) or gas expulsion (rocket firing) is assumed to require environmental interaction, due to conservation of momentum. As first noted in  (Wisdom, 2003) and later in (Gu\'eron, 2009) and (Avron et al, 2006), in curved space or spacetime the non-commutativity of translations permits translation without momentum exchange, just as falling cats and lizards can self-deform to reorient in flat space without environmental interaction. Translation in curved space can occur not only in gravitationally induced curved spacetime (where translation is predicted to be on the order of $10^{-23}$ m per gait cycle) but also in the curved surfaces encountered by locomotors in real-world environments. Here we show that a precision robophysical apparatus consisting of motors driven on curved tracks (and thereby confined to a spherical surface without a solid substrate) can self-propel without environmental momentum exchange (impulse) via shape changes that can generate gauge potentials that manifest as translations. Our system produces shape changes comparable to the environment's inverse curvatures and generates from zero momentum  forward movement of $10^{-1}$ cm per gait cycle even while resisted by weak gravitational and frictional forces. Dissipation via friction eventually arrests the robot but also imbues it with momentum which can be released upon a cessation of shape changes. This work demonstrates how the interaction between environmental curvature, active driving and geometric phases yields rich, exotic phenomena. 
\end{abstract}
\maketitle

Curved surfaces are ubiquitous in physics, biology, engineering and mathematics but are defined by features that defy intuitions derived from flat space. For example, on a spherical surface the square of the hypotenuse is not the sum of the squares of the legs and ``parallel'' lines meet at the poles and the sum of the interior angles of a triangle grows with the triangle's area. Recently, the inability to form periodic crystals on spherical surfaces~\cite{guerra2018freezing,lopez2011frustrated} has given rise to lively dynamics of essential crystalline defects~\cite{zhang2020dynamics,ellis2018curvature}. And, of course, gravitational interactions themselves are derived from the fundamental curvature of four-dimensional spacetime~\cite{misner1973gravitation} leading to explanations of dynamics such as precessing orbit of Mercury and gravitational lensing of light.



\begin{figure}[t]
  \includegraphics[width=0.45\textwidth]{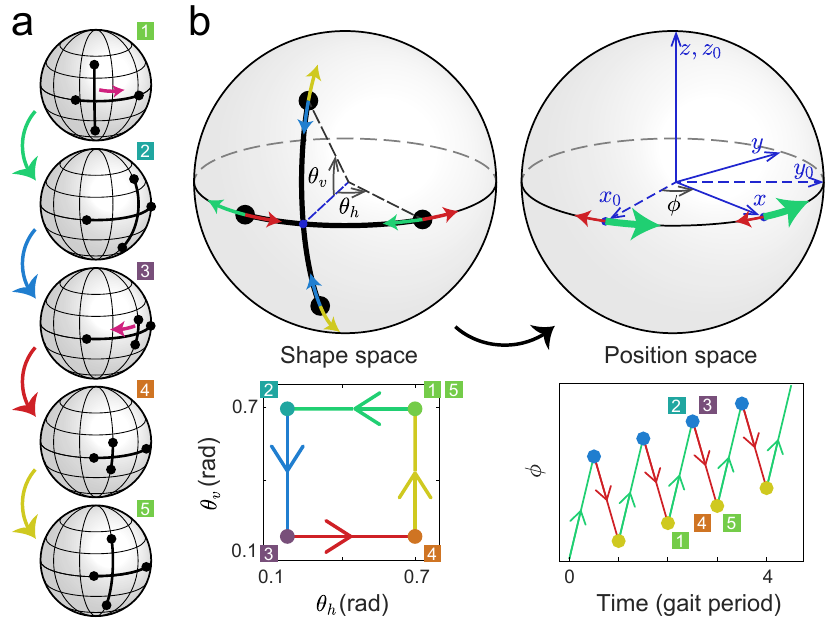}
  \caption{ \textbf{Self-propulsion without reaction forces.} (a) A robot confined to a spherical surface executes a cyclic change of shape to generate net position change. (b) The cyclic change of shape described by the angles $\theta_h, \theta_v$ of motorized weights (black dots) follows an order indicated by the gait diagram below. Due to the variable moment of inertia and the non-commutivity of these operations, this leads to a net change in the robot's position, represented by the rotation $\pos(t)$ applied to the coordinate axes in position space, even in the absence of momentum or external forces. The plot below shows the time evolution of $\phi(t)$ from each stroke of the gait shown in the same color as the gait diagram.
  } 
    \label{fig:setup}
\end{figure}

Less well known is the fact that curved surfaces permit locomotors embedded within them to self propel via translation without exchanging momentum with an environment~\cite{wisdom2003swimming,avron2006swimming,gueron2009adventures} (as is done in swimming, flying and running in typical environments). How can this be?  Consider, in particular,  the prototypical  swimmer confined to (or embedded within) the spherical surface depicted in Fig.~\ref{fig:setup}(a). By propelling masses along the vertical arms, the component of the moment of inertia that relates torque and momentum about the longitudinal axis can be altered, analogous to a process in flat space that somehow altered the mass of an object. By propelling an additional mass along the latitudinal arm on the sphere, angular momentum may be exchanged between this mass and the others during periods in which the robot has different moments of inertia. By pushing itself in one direction when it has low moment of inertia and the other when it has high, the robot may attain a net movement in the first direction, even as the total robot structure maintains zero angular momentum. This is analogous to a falling cat, which instinctively exchanges angular momentum between different parts of its body while contorting itself to alter its moment of inertia.

This process relates to fundamental geometric properties: Whereas a flat plane is invariant under two translations and a rotation, leading to a two-dimensional conserved linear momentum and a one-dimensional conserved angular momentum, a spherical surface is invariant under the three rigid-body rotations of SO(3), leading to a conserved three-dimensional angular momentum. Crucially, these motions, which may be thought of as translations along the sphere, do not commute with each other as do translations in flat space. Consequently, when a robot changes its shape on a sphere as shown in Fig.~\ref{fig:setup}(a), it induces a series of incommensurate motions, so that a closed cycle of shapes induces a net displacement along the sphere, analogous to the rotation achieved by a falling cat.


This behavior is an example of a broader phenomenon in physics, in which dynamically varying patterns can induce a physical transformation known as a \emph{geometric phase}. Geometric phase plays a crucial role in modern physics, from the general-relativistic curvature of spacetime that establishes closed orbits of planets around stars to the Berry curvature that underlies quantum-mechanical effects in graphene, topological insulators and cyclotron motion. And geometric phase even appears in locomotion. As pointed out by Shapere and Wilczek \cite{shapere1989gauge} and developed and applied over the past decades~\cite{marsden1998symmetries,kelly1995geometric,hatton2013geometric,hatton2015nonconservativity,astley2020surprising,chong2021frequency} geometric phase describes how a self-deforming body locomotes in response to drag forces from viscous and frictional fluids to dry friction.


Here we demonstrate experimentally for the first time geometric phase driving dynamics solely induced by the curvature of space, resulting in self-propulsion without environmental force exchange. We do this by converting the abstract picture of the ideal spherical surface swimmer mentioned above to a precision robophysical model. The device's self-propulsion is determined by the geometrical phase induced by its shape changes as it slides motorized masses along curved tracks. Further, under complex dissipative coupling to the environment, this geometric propulsion couples to friction in surprising ways, preventing decay into energy minima and capturing a finite-momentum state in a fixed position.


\section*{Shape change dynamics in curved space} 

\begin{figure}[ht!]
  \centering
  \includegraphics[width=0.37\textwidth]{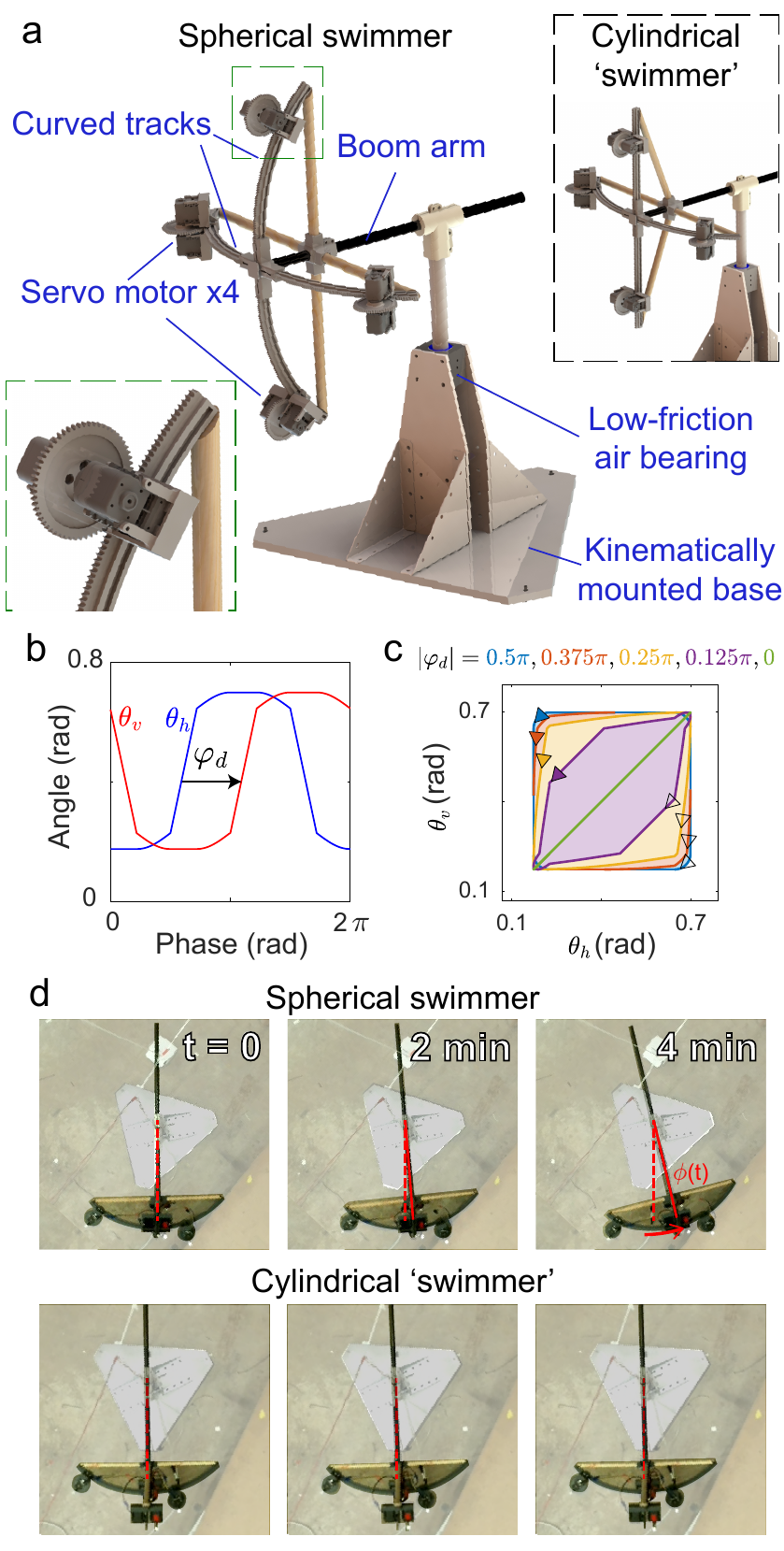}
  \caption{\textbf{Robophysical curved space swimmer.}
  (a) The experimental apparatus, rendered here schematically, confines the robot's active masses to a geometric (but not material) spherical surface of radius 46 cm via 3D-printed curved tracks along which servo motors ($116$ g for each of the four motors) propel masses of themselves and the tracks ($388$ g). The robot is free to rotate about a central pivot with a low-friction air bearing and mounted on a kinematically adjustable base used to minimize the torque induced by gravity. In contrast, the upper-right inset depicts a robot that, due to the straight arms, is confined to a cylindrical surface along which no corresponding motion can be induced via swimming. The lower-left inset shows the coupling between a motor and a curved track. (b) A phase difference $\varphi_d$ between the robot's horizontal and vertical strokes breaks time-reversal and spatial inversion symmetry, as required for forward swimming.
  (c) The displacement of the robot per stroke, in the absence of external forces, is obtained as an integral of geometric phase over the shaded regions enclosed by the gaits in shape space. (d) Chronological snapshots of the spherical swimmer and the cylindrical `swimmer' show the swimming of the former is significant while the latter is vanishingly small. See the red trajectory in Fig. \ref{fig:phase}, \ref{fig:plateau}(a) for the displacement over time and the supplementary movie for the videos of these two experiments.}
  \vspace{-5mm}
  \label{fig:expt}
\end{figure}


Testing the idea that self propulsion can occur in curved environments without forces requires confining the robophysical model to a curved surface while achieving control over its environmental coupling. While experiments often occur in flat planes whose dynamics approximate that of an ideal Euclidean plane, capturing the essence of a sphere is more challenging. In particular, no widely available method exists for placing particles on a solid spherical shell while simultaneously minimizing both the effect of gravity (which would drive particles towards the bottom of the sphere) and friction (which would prevent us from isolating the novel curvature-induced motion from more conventional effects). Instead, we opt for a solution in which we attach the robot to a rigid boom arm free only to rotate about the vertical axis shown in Fig. \ref{fig:expt}(a). 

Masses are robotically propelled along curved tracks whose radius coincides with the length of the boom arm, ensuring that the robot's mass is confined to a spherical surface. The constraint of the boom arm ensures that all forces/torques would move the robot vertically or radially are negated, while freely permitting horizontal direction along the spherical surface. Consequently, the dynamics of the apparatus is well-described by that of an ideal sphere. Notably, the robot's ability to move does not violate the usual rule against movement without forces in three-dimensional space because the full three-dimensional dynamics in fact include normal forces on the robot supplied by the boom arm that confines it to the sphere.

This arrangement is achieved via  precision servo motors connected to gears that move without slipping on the robot's 3D-printed toothed tracks that can be generated with arbitrary curvature profile. The tracks are connected to the  central shaft, which rotates in air bushings with low friction. The base of the system is fixed on the hard ground via kinematic mount, which constrains motion of the base. 







We control the motors' positions on the tracks to prescribe a ``gait'': a closed path in shape space parameterized by the position of the motor positions in horizontal ($\thetahorz$) and vertical ($\thetavert$) directions, as shown in Fig.~\ref{fig:setup}(b). The swimmer is constrained to move angularly by the beam arm [Fig.~\ref{fig:expt}(a)].
This design can minimize environmental forces due to friction and gravity, although we also demonstrate that these couple to the geometric phase to generate additional exotic phenomena.

We  test whether self-deformation in the presence of curvature  can generate locomotion without significant environmental forces. We place the robot on the equator of the sphere and, for simplicity, restrict ourselves to shapes that are symmetrical under reflections across the equator. Such shape changes, combined with the apparatus design, constrain the robot to translate along the equator---i.e. to rotate about the north pole. 
The smooth trajectory through shape space, designed to minimize jerks and maximize motion, is shown in 
Fig.~\ref{fig:setup}(b) and further explicated in Sec.~2 of the SI.

\begin{figure}[t]
  \centering
  \includegraphics[width=0.48\textwidth]{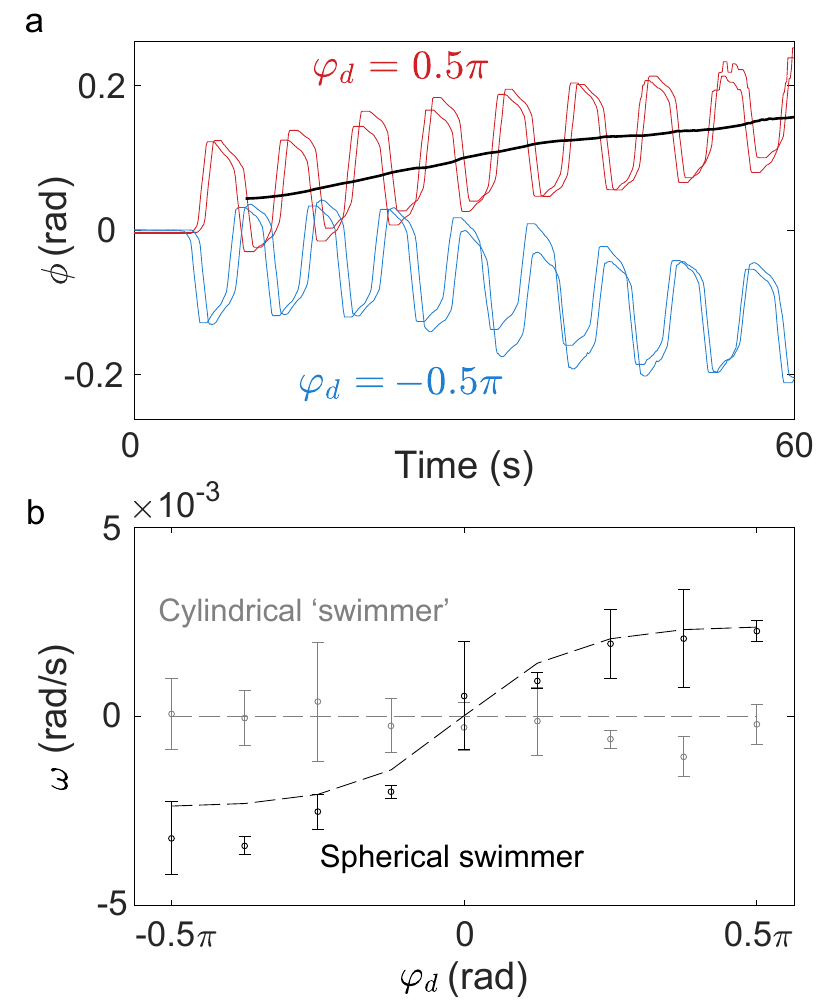}
  \caption{\textbf{Self propulsion via geometric phase}
  (a) The robot, initially at rest, swims forward with an average initial angular velocity $\omega_i=\dot{\bar{\phi}}$ over the course of several strokes. The black thick line shows the time-averaged position $\bar{\phi}$. (b) The observed initial velocities match those predicted from the geometric phase, $\omega_g$, (dashed lines) with variable gait controlling the speed and direction of the robot swimming on the sphere, in contrast to the robot on the cylinder, which cannot achieve significant net movement.
  }
  \vspace{-5mm}
  \label{fig:phase}
\end{figure}

In our primary result, 
as  predicted by Wisdom~\cite{wisdom2003swimming} and indicated in Fig.~\ref{fig:setup}(a), the persistent cycling through shape space shown in Fig.~\ref{fig:expt}(b,c)  leads the robot to translate back and forth, yet the effect of Gaussian (intrinsic) curvature permits a relatively small net motion [Fig.~\ref{fig:phase}(b)] which depends on the direction of the cycle in the configuration space.
And, as predicted, because force-free propulsion relies on this curvature of the doubly-curved sphere, a robot confined to a singly-curved cylindrical surface [inset of Fig. \ref{fig:expt}(a)] does not exhibit this propulsion [Fig.~\ref{fig:phase}(b) and Fig.~\ref{fig:plateau}(a) inset].  Further, in contrast to Wisdom's force-free model, the robotic swimmer's motion saturates at a finite displacement, per Fig.~\ref{fig:plateau}(a)'s main panel.

While we have now realized force-free swimming, as occurs over short times in Fig.~\ref{fig:phase}(a), the robophysical testing reveal more complex phenomena resulting from the interplay between this geometric phase and environmental effects, as in Fig.~\ref{fig:plateau}(a).
We now develop an analytical theory and use numerical simulation to rationalize these results. The full equations of motion described in Sec.~3 of the Supplementary Information  simplify for the equatorial spherical swimmer to the scalar associations  between  the angular velocity, angular momentum and torque about the vertical axis:

\begin{subequations}
\begin{align}
\label{eq:eom}
\dot{\pos}(t) &=  \frac{\angmom(t) - \dot{\inner}(t)}{\moment(t)},
\\
    \dot{\angmom}(t) &= \torque(\pos,\dot{\pos})= -\cf \, \textrm{sgn} \dot{\pos}- \spring \pos.
\end{align}
    \label{eq:torque}
\end{subequations}


The moment of inertia $\moment(t)$  and the internal effective angular momentum $\dot{\inner}(t)$ (the angular momentum induced by the shape change that the robot would have for $\dot{\pos}=0$) both depend on time with period $\period$ via the shape-change gait.
The torques due to friction and gravity are measured, rather than left as free model parameters, in additional experiments as described in Secs.~4,5 of the Supplementary Information.

As a consequence of the dynamics of Eq.~(\ref{eq:torque}), even when angular momentum and torque vanish, the robot's gait causes it to advance at an average angular velocity

\begin{align}
    \label{eq:berryphase}
    \omegagait = \frac{\Delta \pos}{\period} =  -\frac{1}{\period}\int_0^\period \moment^{-1}(t) \dot{\inner}(t) d t = -\frac{1}{\period}\oint \frac{d \inner}{\moment}.
\end{align}

\noindent The final expression reflects the \emph{geometric} nature of this movement as a Berry phase that depends on the  path through shape space but not on the rate at which it is traversed.

\section*{Steady locomotion without force} 

As posited in Fig.~\ref{fig:setup} and shown in Fig.~\ref{fig:phase}(a), in the absence of additional forces, a robot initially at rest would, upon initiating a particular series of shape changes, rotate around the equator of its spherical universe at a rate described by Eq.~(\ref{eq:berryphase}), a behavior analogous to the general-relativistic formulation of Wisdom~\cite{wisdom2003swimming}. Instead, we address the complex coupling between this geometrical phase and the robot's coupling to its environment reflected in the torques of Eq.~(\ref{eq:torque}).

Perhaps surprisingly, while the geometric phase is evaluated via a nonlinear numerical integration and the Coulomb friction is highly nonlinear, the interplay between the two can be treated analytically. This process, shown in the `Two diverging friction models' section of the Methods section is done in the \emph{rotating wave approximation} in which, drawing inspiration from techniques developed in optical physics~\cite{jaynes1963comparison} we decouple the rapid oscillations of the robot from the weaker influences of external torques:

\begin{align}
    \nmoment{0} \ddot{\avepos} &= - \frac{4 \cf \nmoment{0}}{T|\ddot{\inner}|} \left[  \dot{\avepos} - \omegagait \right] - \spring \avepos. \label{eq:eomtwo}\\
    \dot{\angmom} &= - \frac{4 \cf}{T|\ddot{\inner}|} \angmom  - \spring \avepos.\end{align}

\noindent Here, $\avepos(t)$ is the time-averaged position, removing the rapid, high-amplitude oscillations due to the gait motion, as shown in Figs.~\ref{fig:phase}(a),~\ref{fig:plateau}. $\nmoment{0}$ is the inverse of the time-averaging of the inverse moment of inertia. $\ddot{\inner}$ is the rate of change of the internal angular momentum at the time in the gait at which it vanishes. This linear approximation relies on the smallness of the external angular momentum $\angmom$ relative to the internal angular momentum $\dot{\inner}$  For larger angular momenta, higher-order terms become relevant.


 We thus arrive at the effective dynamics of the curvature swimmer. Provided that the external torque does not vary significantly within a single stroke and the system remains in the linear force regime, the system attains an emergent form of the viscously damped linear harmonic oscillation, despite the nonlinearity of the Coulomb friction. The most striking feature is the uniform force field, proportional to both the geometric phase and the Coulomb friction. The periodic shape changes are reminiscent of a Floquet theory, yet the combination of a time-dependent force with a time-dependent inertia permits net forward motion.

The \emph{low-torque} regime occurs when the swimmer, initially at the bottom of a shallow energy well engaged in a neutral swimming motion (which does not lead to self propulsion but maintains continuous self-deformation  to prevent static friction), shifts into a forward swimming motion, as shown in Fig.~\ref{fig:plateau}(a). The robot's motors drive its masses around the spherical surface in the trajectory shown in Fig.~\ref{fig:expt}(b). The parameter $\varphi_d$ controls the offset between the motions of the horizontal and vertical arms, which is crucial to break time-reversal and spatial-inversion symmetries, as is necessary for a swimming gait. Here we refer to a gait with nonzero $\varphi_d$ and thus nonzero swimming as a swimming gait and refer to a gait with zero $\varphi_d$ as a null gait. See the SI movie for demonstrations of the null and swimming gaits, and examples of swimming motion for $\varphi_d = \pm\pi/2$. The geometric phase $\omegagait$ is shown in Fig.~\ref{fig:phase}. Over short periods, the forces effect only a small change in the swimmer's momentum while the swimmer nevertheless advances at a measured initial angular rate $\omega_i=\dot{\bar{\phi}}(0)$. As seen in  Fig.~\ref{fig:phase}(b), the rate predicted (dashed line) by the calculated geometric phase $\omegagait$ (i.e., $\omega_i = \omegagait$) is in agreement with the observed initial rate, providing strong validation of the geometric theory.

\section*{Environmental forces and momentum without locomotion} 

\begin{figure}[t]
  \centering
   \includegraphics[width=0.48\textwidth]{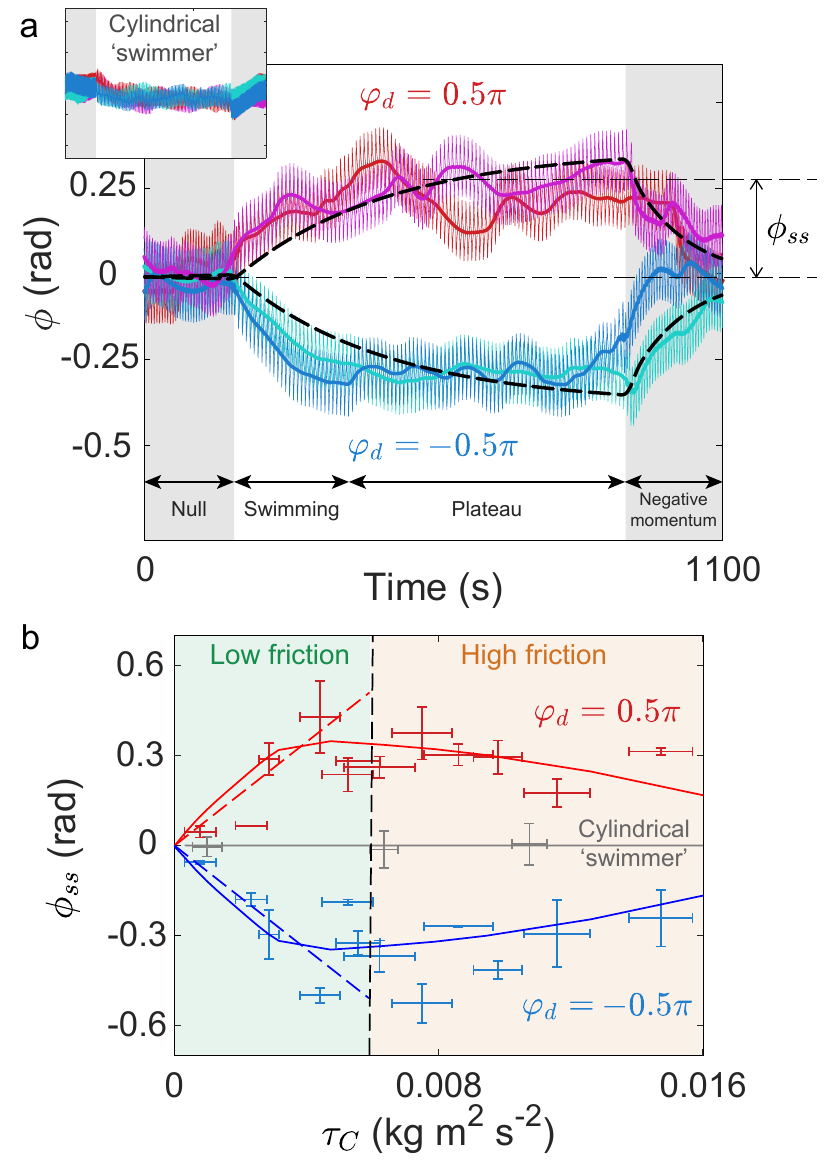}
  \caption{\textbf{Geometric swimming in the presence of environmental effects} (a) Evolution of  $\phi(t)$ of curved swimmer and cylindrical non-swimmer (inset) for Coulomb friction $\cf = 3.4 \times 10^{-3}$ kg m$^2$/s$^2$ and different $\varphi_d$'s. See the SI movie. There are two trials for each $\varphi_d$. (b) The steady-state plateau $\phi_{ss}$ as a function of $\cf$. The dots with error bars, solid lines show the experiment and simulation [Eq.~(\ref{eq:torque})] respectively. To understand how the plateau increases with small friction ($\tau_C < |\ddot{\alpha}|$), we show the first-order theory (Eq.\ref{eq:plateau}) with dashed lines until around the predicted cutoff friction ($\tau_C = |\ddot{\alpha}|$).}
  \vspace{5mm}
  \label{fig:plateau}
\end{figure}

Over longer times, additional vibrations, resonances and nonlinearities preclude quantitative agreement between experiment and theory, yet qualitative agreement is nevertheless observed throughout the trajectory in Fig.~\ref{fig:plateau}(a). These trajectories contain a surprising feature: unlike the typical behavior of dissipative systems in flat space, the curvature drives the swimmer to plateau at a finite offset from the bottom of its potential well. The geometric theory of Eq.~(\ref{eq:eomtwo}) predicts $\tau(\avepos_{ss}) = \frac{4 \cf \nmoment{0}}{\period |\ddot{\inner}|}\omegagait$ and leads to a steady-state plateau of

\begin{align}
\avepos_{ss} = \frac{4 \cf \nmoment{0}}{\spring\period |\ddot{\inner}|}\omegagait.
    \label{eq:plateau}
\end{align}

\noindent This prediction of linear dependence between plateau height and friction strength is observed at low friction strength, as shown in Fig.~\ref{fig:plateau}(b) and levels off at approximately the amount friction strength predicted by theory, which is $|\ddot{\alpha}|$ (see Section 7 of the SI).

\begin{figure}[t]
  \centering
   \includegraphics[width=0.48\textwidth]{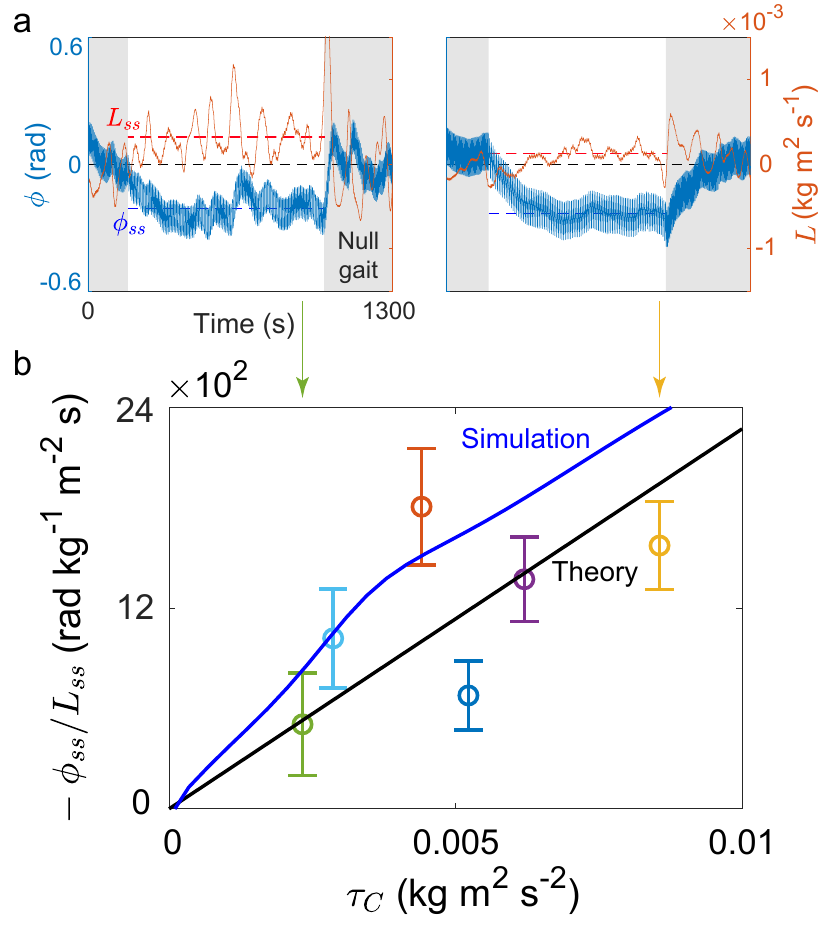}
  \caption{\textbf{Negative momentum in the steady state via environmental effects} (a) Simultaneous evolution of $\phi$ and $L$ for low and high friction ($\tau_C=0.0023, 0.0089$ kg m$^2$ s$^{-2}$) for a swimmer conducting a swimming gait sandwiched by two null gaits. Note that the steady-state plateau $\phi_{ss}$ and angular momentum $L_{ss}$ have opposite signs. (b) The ratio between the $\phi_{ss}$ and $L_{ss}$ for various torques of friction $\tau_C$. The black line shows the theory $-\phi_{ss}/L_{ss}=4\tau_C/(T|\ddot{\alpha}|\tau_g)$ with all parameters measured from experiments. The blue line shows the simulation result.}
  \vspace{-5mm}
  \label{fig:negMomentum}
\end{figure}

The trajectory in Fig.~\ref{fig:plateau}(a) naively suggests that the swimmer begins with a finite momentum that falls to zero as its position plateaues.
 As our analysis reveals, the reverse is true: at the beginning of the swimmer's journey, when its velocity is greatest, it lacks momentum. In contrast, once its average velocity in one direction vanishes, the momentum is maximal and points in the opposite direction. Because the swimmer advances without momentum the dissipative forces that arrest the swimmer's forward progress also impart an impulse that leaves it with nonzero momentum. To illustrate this, in Fig.~\ref{fig:plateau}(a), Fig.~\ref{fig:negMomentum}, we suspend the swimmer's forward stroke, replacing it with a null gait to prevent static friction. At this moment, normal classical physics, in which momentum and velocity are in proportion to one another, reasserts itself and the swimmer's negative momentum causes it to swing backwards towards its origin. We emphasize that this is not due to external forces: gravity is weak and friction opposes this motion. This is due purely to the momentum concealed by and compensated for in the swimmer's gait. The curvature swimmer thus displays the exotic behavior of using dissipative (frictional) forces to \emph{increase} the magnitude of its momentum in the rest frame of its environment. This relationship between steady-state displacement and momentum remains in qualitative agreement with simulation and theory (without free parameters) as friction is varied.

\section*{Conclusions}\color{black}
In summary, we have experimentally realized and theoretically characterized the movement of a robot through a curved (spherical) space without relying on any momentum or reliance on environmental forces to translate, in contrast with all other observed systems. We demonstrated that this purely geometric effect couples to both conservative and dissipative forces present in real environments. In particular, we have shown how coupling between Coulomb friction and the geometric phase generates an effective force on the robot that imparts an impulse that reduces its velocity while increasing its momentum in the opposite direction. This behavior sheds light both on Wisdom's proposed locomotion via spacetime curvature~\cite{wisdom2003swimming} and, more immediately, on a geometric effect always present when robots move on curved surfaces.
As shown here, this effect can become dominant when the robot's body is comparable to the inverse curvature of the surface, which appears to extend to gravitational curvature of spacetime as well~\cite{wisdom2003swimming}.
Further, the apparatus presented here can function as a test-bed for additional exotic behavior on curved surfaces, related to more complex and variable swimming gaits, nonlinear effects and collective behavior.

\section{Acknowledgments} We thank Baxi Chong, Andras Karsai, Bo Lin, Jack Wisdom and Lutian Zhao for advice and discussion. Funding for D.I.G, D.Z.R, S.L, T.W, J.M, V.H.K, Y.O.A. and E.A provided by the Army Research Office under contract W911NF-19-1-0056; funding also provided to D.I.G. by a Dunn Family Professorship.
\bibliography{swimmer_references}
\section{Author information}

Experiments were performed by S.L, T.W, V.K, Y.A, E.A and D.I.G. Simulations were performed by S.L, T.W and J. M. Analytic theory was developed by S.L, J.M, D.I.G and D.Z.R. The paper was conceived by D.I.G, J.M and D.Z.R. The paper was written by S.L, T.W, V.K, D.I.G and D.Z.R.

\appendix

\section{Methods}

\section{Experiment apparatus}
The swimmer is composed of four controlled motor modules functioning as moving masses on four curved tracks 3D-printed with PLA (polylactic acid) filament. The four tracks as shown in Fig. \ref{fig:setup}(b) of the main text are mounted to a horizontal carbon fiber beam arm. This arm is attached to a vertical steel shaft, which can rotate about the z axis. To create a low friction environment for the swimmer, the steel shaft is supported by two air bearings and one air bushing.

To reduce the residual gravity from any small incline of the swimmer, the apparatus is attached to a kinematically mounted base. To allow varying friction to impede swimmer motion and probe environmental interaction effects, we attach a block of polyurethane foam to a screw-adjusted sliding block, allowing adjustable normal contact between the shaft and the foam. 

Motion of servo motors (Dynamixel AX-12A, Robotis) is controlled by position commands. The instantaneous motion of the swimmer and the positions of individual motors over time are recorded by an optical tracking system.

For more details of the apparatus, please see the Section 1 of the Supplementary Information.

\section{Numerical integration}

We implement the same gait as the commanded shape change sent to the motors by interpolating these discrete signals in the differential equations. To lower the computation cost of interpolation, we use a numerical scheme with fixed steps (forward Euler). The test of convergence with step size $h$ shows a global error (i.e., the error in the position $\phi$) of $O(h)$ (and therefore local error of $O(h^2)$) as expected for a first-order scheme. We use the step size $h=3.1\times 10^{-4}$ s such that the relative error is $1.8~\%$. See Section 6 of the SI for the details.

\section{Shape dynamics}

As described in the main text, we consider a robot undergoing shape changes while confined to a spherical surface of radius $\rad$. Here, we describe some of the technical details of the general formulation.

Let our robot's shape $\shape(t)$ be described by masses $\mass_i$ at points $R \pointi(t)$. We choose units such that we can set the radius to one. In the case of a continuous robot we would need to define some other sets of degrees of freedom to parameterize the shape of the robot. From the general definition that angular momentum is the cross product of position with momentum, the shape change alone induces an angular momentum:

\begin{align}
    \sum_i \mass_i \pointi \times \veli.
\end{align}

\noindent However, the dynamics will also select some rigid rotation of the robot, $\rot(t)$. This will not only rotate the above contribution but generate a new term, as the position of mass $i$ is $\rot(t) \pointi(t)$ and its velocity is $\dot{\rot} \pointi + \rot \veli$. Hence, the angular momentum is:

\begin{align}
    \angmomvec &= \sum_i \mass_i (\rot \pointi)\times (\dot{\rot} \pointi + \rot \veli) 
    \\
    &= \sum_i \mass_i (\rot \pointi) \times \rot \left(\rot^{-1} \dot{\rot} \pointi + \veli\right)
 \\
&= \rot \sum_i \mass_i \pointi \times \left(\rot^{-1} \dot{\rot} \pointi + \veli\right).
\end{align}

We recognize the term $\rot^{-1} \dot{\rot}$ as an element of a Lie algebra. It is the skew-symmetric generator of infinitesimal rotations. When the angular momentum in rotated coordinates, the shape and the shape change are all known, this generator may always be solved for. One may then integrate these generators numerically to obtain the total rotation induced by a set of shape changes. As shown by Shapere and Wilczek for general rigid-body rotations~\cite{shapere1989gauge}, in the limit of small shape changes (e.g., a robot small compared to the radius of curvature) this may be expressed as a gauge theory. 

However, in order to achieve a more experimentally accessible regime, we must impose large shape changes, in which the robot evolves over length-scales comparable to the inverse curvature (i.e., radius) of the space in which it lies. In order to generate an analytically tractable theory, we then restrict ourselves to shapes that are symmetrical under reflections about the equator. Consequently, the final term in the angular momentum expression above must point in the $z$-direction. It also follows that the linear operator acting on the generator of rotations is symmetric about this reflection as well. Therefore, in the absence of external forces to the contrary (our robot is fixed to the equator, preventing gravity from breaking this symmetry and drawing it to the south pole) for a robot initially at zero angular momentum all rotations must be about the vertical axis. This leads to the following expression for the remaining component of angular momentum:

\begin{align}
    \angmom \hat{z} &=\left[ \sum_i \mass_i \pointi \times \left(  \hat{z} \times \pointi\right)\right] \dot{\pos}(t) +\sum_i \mass_i \pointi \times \veli, 
    \\
    L &\equiv I(t) \dot{\pos}(t) + \dot{\inner}(t)
\end{align}

The coefficient of $\dot{\pos}(t)$ is the $zz$ component of the moment of inertia tensor, which is denoted $\moment(t)$. The final term, which we denote $\dot{\alpha}$, is the angular momentum the robot would have due to its shape change even if it were not rotating. For example, if the base of the robot is fixed, but one of its legs is proceeding counter-clockwise, it would have counter-clockwise angular momentum.

\section{Calculating Berry phases and applying the rotating-wave approximation}

As discussed in the previous section and the main text, we have an equation of motion describing how the angular momentum and internal shape changes induce changes in the robot's position

\begin{align}
\label{eq:inner}
\dot{\pos}(t) =  \frac{\angmom(t) - \dot{\inner}(t)}{\moment(t)}
\end{align}

\noindent and one governing how external forces modify the angular momentum:

\begin{align}
    \dot{\angmom} = \torque(\pos,\dot{\pos}).
\end{align}

Here, we discuss how an approximate theory emerges under mild assumptions that are well-satisfied in practice. We assume that the angular momentum is varying fairly slowly relative to the timescale of the robot's gait, so that we might expand it as $\angmom(t) \approx \angmom(t_0) + (t-t_0) \dot{\angmom}(t_0) + \ldots$. We also assume that the moment of inertia is time-symmetric about $t=0$, though note that this does not imply that the gait as a whole is time-symmetric (a fully time-symmetric gait could not advance in any direction, at least barring spontaneous symmetry breaking). Now, let us define the $n^{\textrm{th}}$ moment as

\begin{align}
\nmoment{n}^{-1} \equiv \period^{-(n+1)} \int_{-\period/2}^{\period/2} dt \, t^n I(t)^{-1}.
\end{align}

\noindent In these terms, using the fact that odd moments vanish by symmetry, we have that over one period, the time-averaged change in angular position is

\begin{align}
\label{eq:phibardot}
 \dot{\avepos} = - \frac{1}{\period} \oint \frac{ d \inner}{\moment} + \frac{\angmom}{\nmoment{0}} + \frac{\period^2}{2 \nmoment{2}}  \ddot{L} + O(T^4).
\end{align}

That is, to good approximation the advancement of the robot is given by the geometric phase plus the term one would expect in the absence of shape changes: the angular momentum  divided by a time-averaging of the moment of inertia. Corrections to this picture emerge as the angular moment  changes rapidly over a single stroke of the gait. 

Even ignoring these higher-order terms, the portion of the angular position that is not time-averaged may have a substantial amplitude, as indeed is observed in our experimental system. 

\section{Two diverging friction models}

Consider a simple model of viscous friction and a conservative force:

\begin{align}
    \dot{\angmom} = - \vf \dot{\pos} + \tau(\phi).
\end{align}

This time-averages trivially to

\begin{align}
    \dot{\angmom} \approx - \vf \dot{\avepos} + \tau(\avepos).
\end{align}

By combining this expression with the dominant terms in Eq.~(\ref{eq:phibardot}) (after time-differentiating), we have:

\begin{align}
   \nmoment{0} \ddot{\avepos}= - \vf \dot{\avepos} + \tau(\avepos).
\end{align}

These dynamics now resemble those of typical oscillator motion. Indeed, the geometric phase has vanished. Its relevance comes in returning to Eq.~(\ref{eq:phibardot}), in which we see that for a system that is initially not swimming and at rest (angular momentum zero), the initiation of the swimming motion leads to an \emph{initial velocity} proportional to the geometric phase which appears as the first term on the right-hand side of Eq.~(\ref{eq:phibardot}). In the absence of any force, the swimmer would persist at this speed, consistent with Wisdom's picture~\cite{wisdom2003swimming}. When viscous friction is included without force, we see that we have trajectories 

\begin{align}
\avepos(t) = \avepos(0) + \frac{\nmoment{0}}{\eta}\left[ -\frac{1}{\period} \oint \frac{ d \inner}{\moment}  \right] \left[1-\exp\left(-\frac{\vf}{\nmoment{0}} t\right)\right].
\end{align}

At steady state, then, we have:

\begin{align}
\Delta \phi_{\textrm{ss}} &= \frac{\nmoment{0}}{\eta}\left[ -\frac{1}{\period} \oint \frac{ d \inner}{\moment}  \right], \\
\dot{\avepos}_{\textrm{ss}} &= 0,\\
\angmom_{\textrm{ss}} &= \frac{\nmoment{0}}{\period} \oint \frac{ d \inner}{\moment} .
\end{align}

\noindent That is, initially the system has a velocity, given by the geometrical phase, with no momentum. Eventually, its velocity vanishes but its momentum is nonzero, and points in the opposite direction from its previous velocity.

If we include a conservative force, steady state instead requires that the conservative force vanishes, returning the swimmer to the bottom of the potential well. Thus, its geometric phase represents only a temporary escape, before it reaches the same fate as a conventional particle, driven to the bottom of a potential well by dissipative forces.

We turn, then, to a different friction model, one which couples more deeply to the geometric phase and which is more relevant to the experimental apparatus: \emph{Coulomb friction}:

\begin{align}
    \dot{\angmom} = - \cf \textrm{sgn} \dot{\pos} + \tau(\phi).
    \end{align}

\noindent Here, time-averaging leads to a different result. Our original gait, by symmetry, has velocity forward as often as backwards, with the swimmer advancing because it moves forward more quickly than it moves backwards. Thus, any forces resulting from the Coulomb friction come from the potential for angular momentum to change the fraction of the time it spends going forward. The average value of the Coulomb friction term is, then,

\begin{align}
    \dot{\angmom} \approx - \frac{4 \cf}{T} \left(\dot{\inner}^{-1}(\angmom)-\dot{\inner}^{-1}(0)\right) + \tau(\avepos).
\end{align}

Here, the term $\dot{\inner}^{-1}(\angmom)$ appears because, per Eq.~(\ref{eq:inner}), the periods in which the velocity is positive end and beginning with periods in which $\dot{\inner(t)} = \angmom$. The factor of four occurs because this effect at the beginning of such a period is mirrored, by symmetry, with the opposite effect at the end of the period, and because as the period in which this term is positive increases the period in which it is negative decreases.

Because $\dot{\inner}(t)$ is odd about one of its zeros, in expanding the above equation we do not obtain $O(\angmom^2)$ terms. Higher-order terms can be neglected, leading to

\begin{align}
    \dot{\angmom} \approx - \frac{4 \cf}{T|\ddot{\inner}|} \angmom + \tau(\avepos).
\end{align}

Here, $\ddot{\inner}$ is evaluated at one of the zeroes of $\dot{\inner}$. The result is something resembling again viscous friction. However, for our system the distinction between a term proportional to angular momentum and one proportional to angular velocity is profound. 

Indeed, upon combining this equation with Eq.~(\ref{eq:phibardot}) and, again, suitable time derivatives, we obtain the equations of motion for the Coulomb friction case:

\begin{align}
    \nmoment{0} \ddot{\avepos} = - \frac{4 \cf \nmoment{0}}{T|\ddot{\inner}|} \left[  \dot{\avepos} + \frac{1}{\period} \oint \frac{d \inner}{\moment}\right] + \tau(\avepos).
\end{align}

That is, in the time-averaged dynamics, the Coulomb model leads to a viscous-like term, but also to a uniform forcing term from the interaction between the geometric phase and the Coulomb friction. This leads to strikingly different dynamics than in the viscous case. 

In particular, in the absence of the conservative force, the Coulomb curvature swimmer advances forever, reaching a steady-state in which angular momentum vanishes but velocity is given by the geometric phase: 

\begin{align}
\dot{\avepos}_{\textrm{ss}} = - \frac{1}{\period} \oint \frac{d\inner}{\moment}.
\end{align}

\noindent Moreover, in the presence of a conservative potential well, rather than falling to the bottom, the swimmer is able to reach a steady state in which the interaction between the Coulomb potential and the geometric phase allows the swimmer to resist the conservative force:

\begin{align}
    \tau(\avepos_{\textrm{ss}}) = \frac{4 \cf \nmoment{0}}{T^2|\ddot{\inner}|} \oint \frac{d \alpha}{\moment}
\end{align}

\renewcommand{\thefigure}{S\arabic{figure}}
\setcounter{figure}{0}
\section{Supplementary Information}
\section*{S1. Apparatus and experiment setup}

\begin{figure}[t]
  \includegraphics[width=0.45\textwidth]{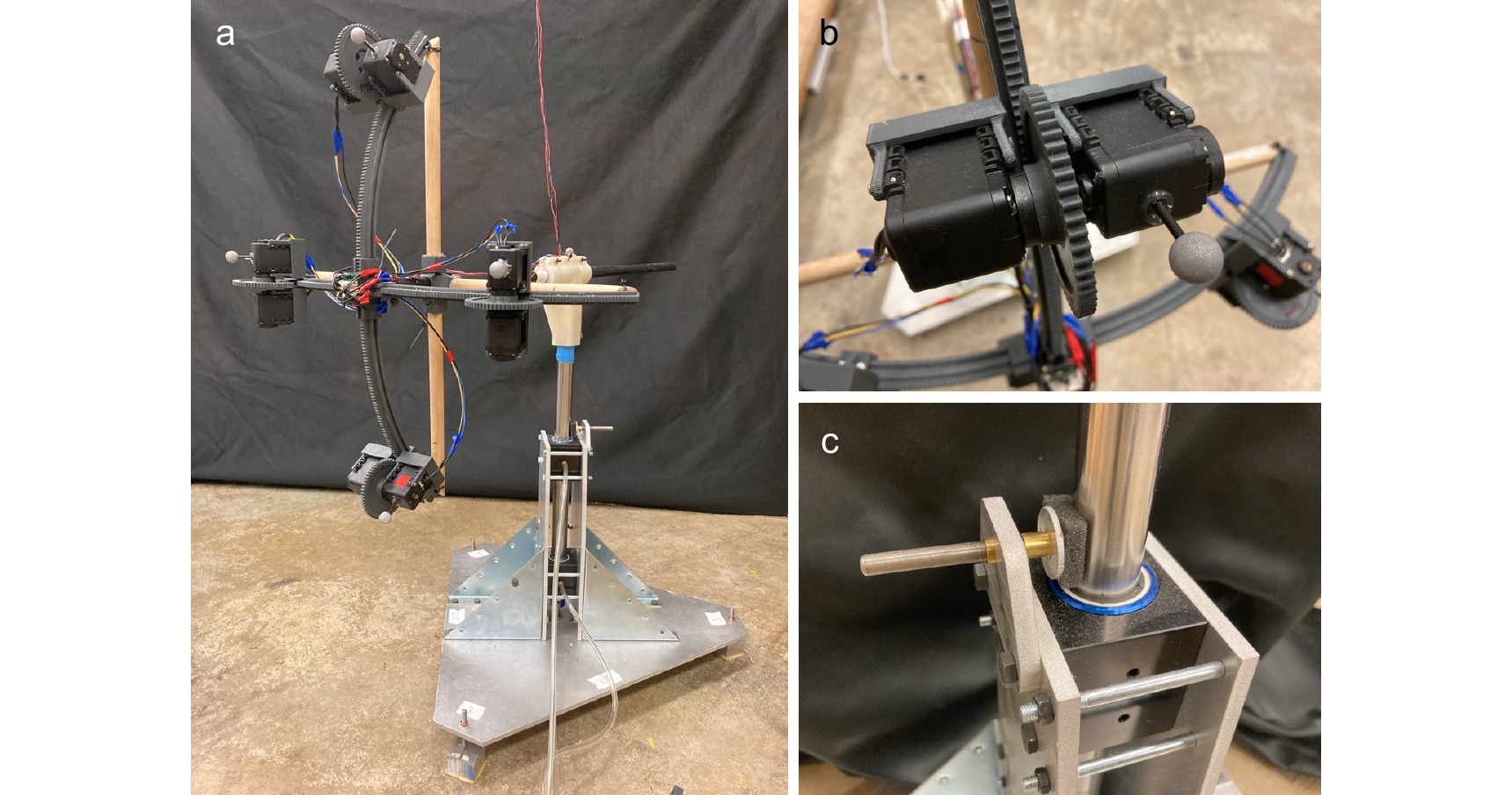}
  \caption{Apparatus and experiment setup of the swimmer. (a) The swimmer. (b) The motor module. (c) The tunable friction block.}
\label{fig:apparatus}
\end{figure}

The apparatus of the robotic swimmer device is shown in Fig.~\ref{fig:apparatus}a. The swimmer rests on a set of three kinematic mounts. Each mount is a 25mm $\times$ 25mm $\times$ 25mm aluminum cube that was CNC machined at Georgia Tech’s Montgomery Machining Mall using a Dayton Mill Drill Machine. One mount has a ball and socket groove (constrains 3 degrees of freedom for motion), another has a grooved slot (constrains 2 degrees of freedom for motion), and the last has a smooth surface (constrains 1 degree of freedom for motion). These mounts were fixed to the ground with super glue. Having the three kinematic mounts ensures that the device has all six degrees of freedom for motion constrained, preventing slipping or wobbling. 

The triangular base of the swimmer was machined out of 0.25 inch aluminum using a Maxiem CNC Waterjet at Georgia Tech’s Flowers Invention Studio. The base is an equilateral triangle with sides that are 550 mm long. The base also has 3 holes (one at each corner) that have a thread brass insert inside. A 0.25-inch-100-thread ultra-high precision set screw is placed inside each brass insert, and the tip of each set screw rests on one of the kinematic mounts. These set screws can be adjusted to make minute changes in the angle of the base. 

The side panels are machined the same way as the base and are attached to the base using M5 bolts and corner brackets so that they stand up vertically. In between the side panels, there are two 25 mm diameter air bushings secured in the mounting boxes. When connected to air, the air bushings blow tiny streams of air radially inward so that any rod inside of them is left ``floating'' inside with minimal friction. Inside the air bushings, there is a 25 mm diameter steel shaft that is 450 mm tall. The bottom of the rod rests on a 25 mm diameter flat air bearing mounted to the center of the triangular base. Using the combination of two bushings and a bearing, the steel shaft is supported vertically with minimal friction and allowed to rotate freely. The air bushings, air bearings, and mounting boxes were from New Way Air Bearings.

On top of the steel shaft is a T-Shaped connector that was 3D-printed in PLA (polylactic acid) plastic using an Ultimaker S3 3D Printer at Georgia Tech’s FLower’s Invention Studio. The connector has two holes that are perpendicular to one another. The bottom hole is rigidly fastened to the steel shaft using nuts and bolts. 
In the other hole of the Connector there is a 15 mm diameter carbon fiber arm, that extends out 400 mm, which stays perpendicular to the steel shaft and parallel to the ground. At the end of this carbon fiber arm are the tracks and motors of the swimmer. 

Each curved track has a center of curvature at the point where the steel shaft and carbon fiber arm meet and a radius of curvature of 400 mm (same as the length the carbon fiber arm extends out). Each has an arc length of 280 mm (forms an angle of 0.7 rad). The four curved tracks are all connected in the middle of the rod on a connector piece. Each track has teeth running along the outside. Traveling on each track is a motor module (Fig.~\ref{fig:apparatus}b). The module features a mount that the AX-12A servo motor (from ROBOTIS) is rigidly attached to using screws and a gear that is rigidly attached to the motor. The gear's teeth line up with the teeth on the track allowing the motor to travel up and down the curve. Opposite  each AX-12A servo motor is a counterweight with a weight equivalent to the AX-12A servo motor that moves along the track with the motor. The four curved tracks, the connector, the four gears, and the four motor mounts were printed in ABS (acrylonitrile butadiene styrene) plastic using a Stratasys uPrint SE Plus Printer.
To support the tracks there are four 15 mm diameter wooden dowel rods that extend from the carbon fiber arm to the end of each curved track ($\sim$25 mm tall). Wooden rods stabilize the tracks and prevent unwanted shaking. 

The four AX-12A servo motors are all connected to a USB communication converter U2D2 (from ROBOTIS) using long soft wires so that the U2D2 does not rest on the swimmer. The wires remain loose and do not interfere with the swimming motion. The servo motors are controlled by position commands via TTL Communication.

A polyurethane foam is attached to the shaft to create tunable friction on the swimmer. The foam is fixed on a sliding block (Fig.~\ref{fig:apparatus}c), which allows us to adjust the normal force, therefore the friction, by adjusting the set screw.

To track the motion of the swimmer in its position space,  six IR reflective markers were attached and tracked the trajectories of the markers using an OptiTrack motion capture system with six OptiTrack Flex 13 cameras. Four markers were attached to the motor modules, one was attached to the pivot position on the T-connector, and the other one is attached on the carbon fiber arm. Real-time 3D positions of the markers were captured at a 120 FPS frame rate. 

\section*{S2. Gait execution}
\begin{figure}[ht!]
  \includegraphics[width=0.45\textwidth]{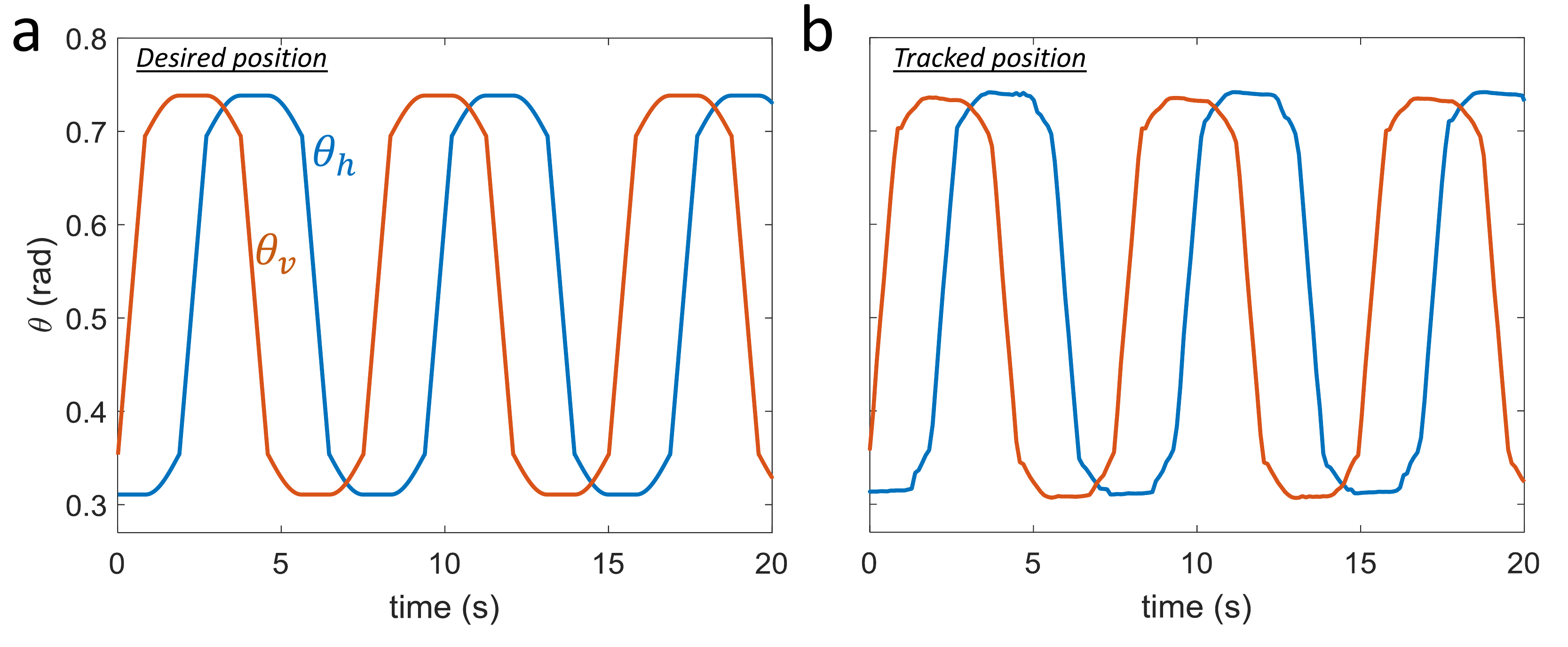}
  \caption{Commanded positions (a) sent to servo motors and the tracked motor positions (b).}
    \label{fig:gaitExpt}
\end{figure}
To avoid mechanical instability, we smooth the connection between the piecewise strokes with sinusoidal functions. Fig.~\ref{fig:gaitExpt}a shows an example of the desired $\theta_v$ and $\theta_h$ for the gait with $\varphi_d = -\pi/2$ positions. In the experiment, the rotary motor was controlled by the angular positions of its wheel gear. Since the relation between $\theta_v$ ($\theta_h$) and the position of a vertical (horizontal) motor on the track is linear, and the motor position on the track is proportional to the angular position of the motor wheel gear, the position commands to motors were derived from $\theta_v$ and $\theta_h$ by a simple linear mapping. The swimmer executes the gait when all motors follow the sequences of position commands. Fig.~\ref{fig:gaitExpt}b shows the $\theta_v$ and $\theta_h$ observed from the motion tracking system to show the quality of gait execution.

\section*{S3. Equation of motion}

The total angular momentum of the swimmer is composed of three parts: the contribution from the vertical motors, the horizontal motors, and the track system supporting the motors together with the supporting rod.
\begin{eqnarray}
L&=&L_{\text{vertical}}+L_{\text{horizontal}}+L_{\text{track}}\\
&=&2m_vR^2\cos^2{\theta_v}\dot{\phi}+2m_hR^2(\dot{\theta}_h+\dot{\phi})+I_{\text{track}}\dot{\phi}
\end{eqnarray}

If we collect the terms with $\dot{\phi}$ together, we have (following the ``shape dynamics'' section of the Methods)

\begin{eqnarray}
L=I\dot{\phi}+\dot{\alpha}\label{eq:angMom}
\end{eqnarray}

where now

\begin{eqnarray}
I(t)&=&2m_v R^2 \cos^2{\theta_v(t)}+2m_h R^2+I_{track}R^2\label{eq:I},\\
\alpha(t)&=&2m_h R^2 \theta_h(t).\label{eq:alpha}
\end{eqnarray}

The torque that changes the angular momentum is composed of two parts, being the contribution from the slight residual gravity and Coulomb friction.
\begin{eqnarray}
\frac{dL}{dt}=\tau=A_g+A_C\label{eq:EoM}
\end{eqnarray}
The force from the residual gravity is caused by the mass of the swimmer on the equator, which normal slightly misaligns with the direction of Earth's gravity with an angle of $\theta_g$. The residual gravity potential contributed by the two horizontal motors, two vertical motors, and the track compose a total residual potential energy of $V=-(2m_v+2m_h+m_{\text{track}})gR\sin{\theta_g}\cos{(\phi-\phi_0)}$. This leads to a torque of
\begin{eqnarray}
A_g&=&-\partial V/\partial\phi\\
&=&-(2m_v+2m_h+m_{\text{track}})gR\sin{\theta_g}\sin{(\phi-\phi_0)}
\end{eqnarray}
where $\phi_0$ is the azimuthal position with the lowest potential energy. Without  loss of generality, we set $\phi_0=0$ so the torque from gravity and assume $\phi$ is small and finally arrive at $A_g=-\tau_g\theta$ where $\tau_g=m_s gR\sin{\theta_g}$ where $m_s=2m_v+2m_h+m_{\text{track}}$.

\begin{figure}[ht!]
  \includegraphics[width=0.4\textwidth]{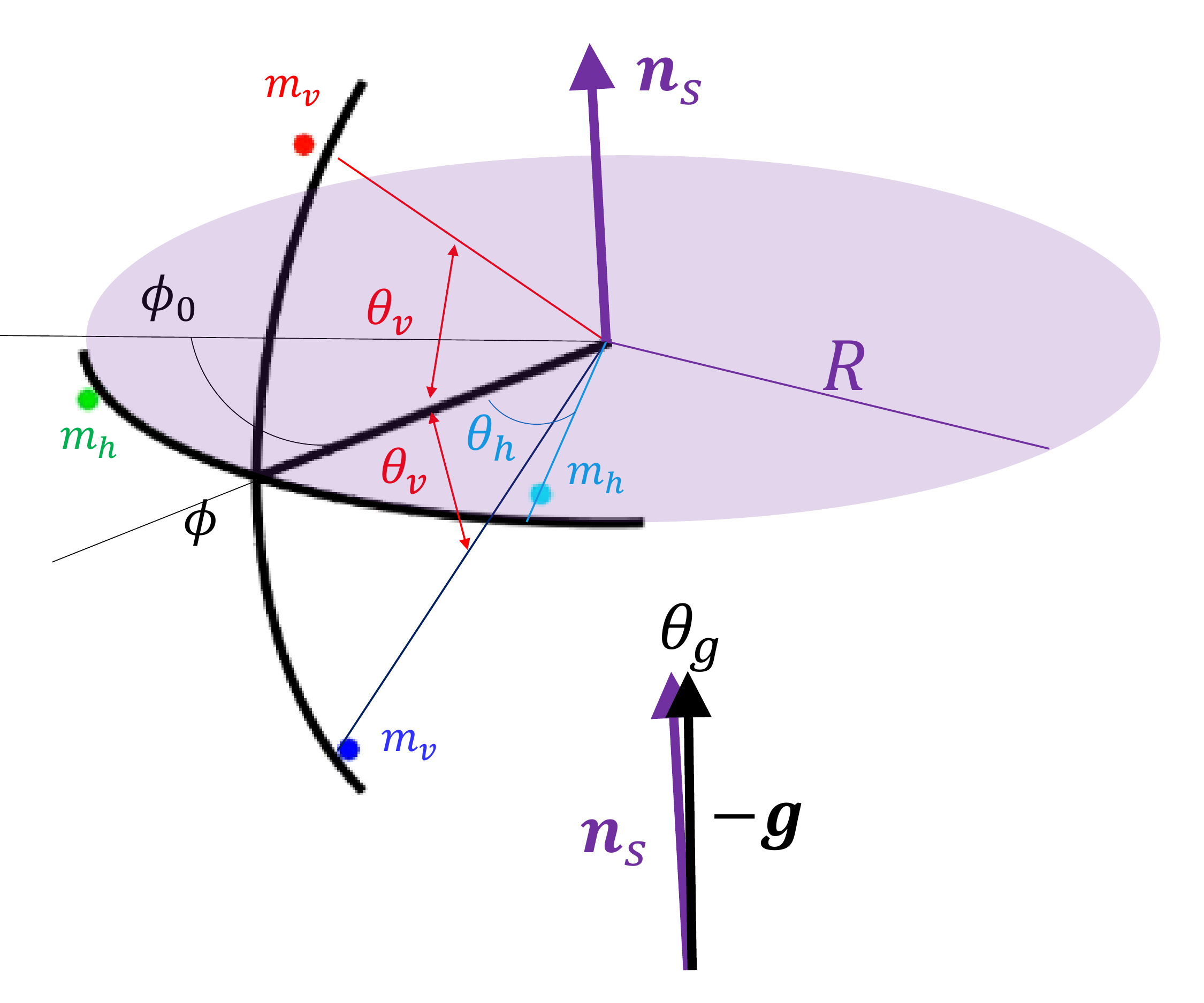}
  \caption{\textbf{Setup of the swimmer.} The swimmer with two horizontal motors (each with mass $m_h$) and two vertical motors (each with mass $m_v$) rotates about $\mathbf{n}_s$, the normal of the equator shown as the light purple plate. The position of the swimmer, $\phi$, the azimuthal angle of the beam arm connecting the center and the curved track arms, evolves as the positions of the motors ($\theta_h,\theta_v$) move. Ideally, the normal of the equator $\mathbf{n}_s$ should be aligned with the opposite direction of gravity $-\mathbf{g}$. In realistic experiment, we characterize the small residual gravity by the angle $\theta_g$ ($\sim10^{-4}$ rad) between $\mathbf{n}_s$ and $-\mathbf{g}$. We denote the minimal position of the gravity potential as $\phi_0$.}
\label{fig:swimmerSetup}
\end{figure}

The torque from friction has a constant magnitude $\tau_C$ and a direction opposite to the angular velocity so

\begin{eqnarray}
M_C = -\tau_C \text{sgn}(\dot{\phi})
\end{eqnarray}

Piecing all above, we have the equation of motion as
\begin{eqnarray}
L&=&I\dot{\phi}+\dot{\alpha}\\
\dot{L}&=&-\tau_C \text{sgn}\dot{\phi}-\tau_g \phi\label{eq:EoM}
\end{eqnarray}

where $\tau_g=(2m_v+2m_h+m_{\text{track}})gR\sin{\theta_g}$.

\section*{S4. Friction characterization}
\begin{figure}[ht]
  \centering
  \includegraphics[width=0.45\textwidth]{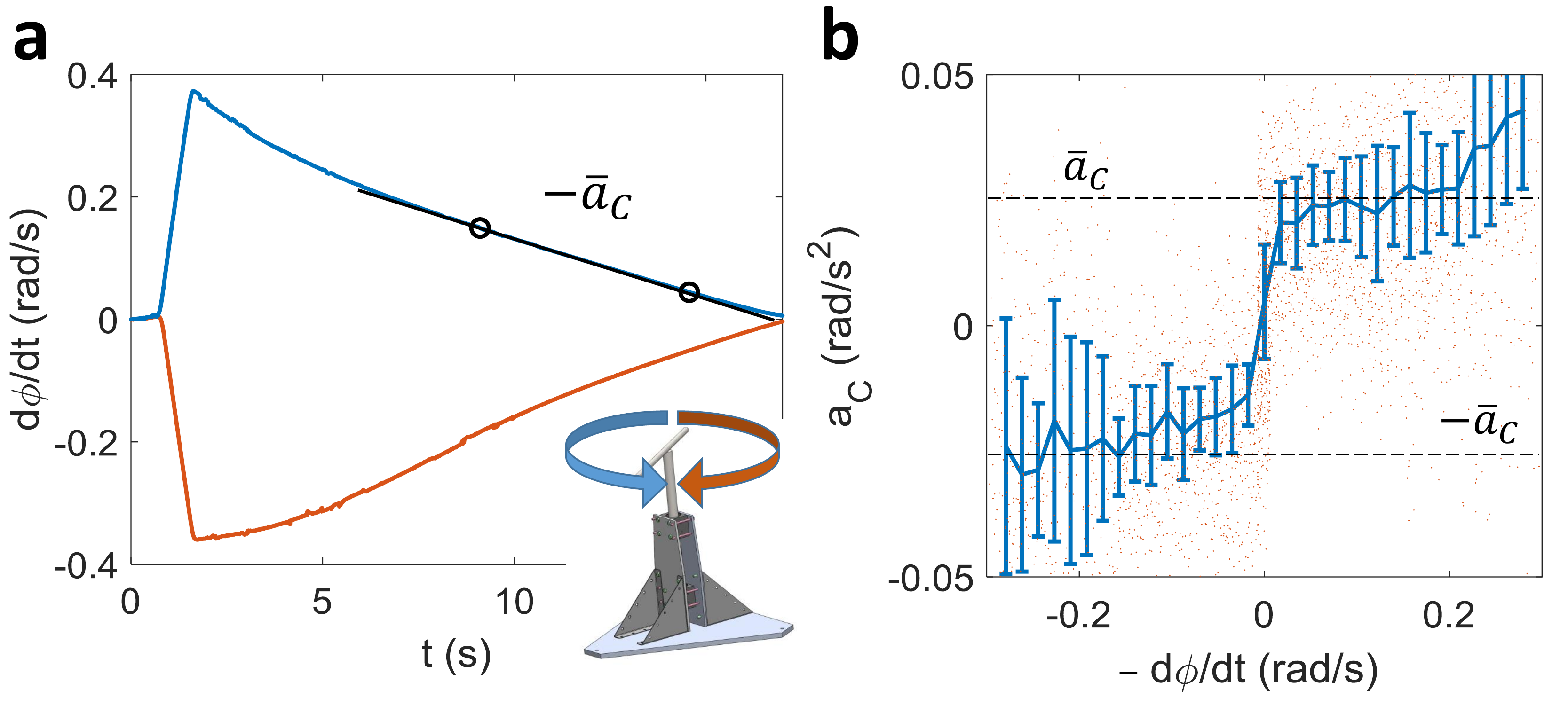}
  \caption{(a) The decay of angular velocity $\dot{\phi}$ over time in experiment. (b) Angular acceleration over $\dot{\phi}$. The orange dots show the raw data. The blue error bars show the median and the middle quartiles of the binned data.}
    \label{fig:fricChar}
\end{figure}

To characterize and measure the friction, we tracked the decay rate of angular velocity as shown in Fig.\ref{fig:fricChar}a. Fig.\ref{fig:fricChar}b, an example of such experiment, shows the instantaneous decay rate $a_C$ as a function of the angular velocity $\dot{\phi}$. We note that it is reasonably anti-symmetric about zero, giving a symmetric friction status such that the possible ratchet effect, which could introduce unwanted swimming, is nominal. When reporting the friction in the main text, we use the acceleration evaluated for the range of angular velocity between $0.175$ rad/s and $0.035$ rad/s, which is the typical range of angular velocity in the swimmer's experiments. To average out the possible slight gravity residue effect, we performed experiments at $5$ different azimuthal positions evenly spaced in $(0,2\pi)$. To avoid the ratchet effect, we only performed swimmer experiments when the discrepancy of friction between the clockwise and counterclockwise value is less than $10~\%$.

To convert the acceleration to the torque from friction $\tau_C$, we multiply the average acceleration magnitude $\bar{a}_C$ by the total moment of inertia of the swimmer $I_0=m_sR^2$ where $m_s$, $R$ are the mass and the radius of the swimmer, respectively.

\section*{S5. Inferring the slight gravity potential}

Since the gravity potential well from the slight tilting of the equator is so shallow that direct measurement from devices such as bubble meter or optic tracking does not have sufficient resolution due to possible error from the mounting of trackers, we infer its depth using the long oscillation period resulted from it.

Particularly, we perform a very long null gait in which only the horizontal motor moves. The averaged $\phi$ (i.e., the envelope $\bar{\phi}$) is given by Eq. (4) in the main text as
\begin{eqnarray}
\nmoment{0} \ddot{\avepos} &= - \frac{4 \cf \nmoment{0}}{T|\ddot{\inner}|} \left[  \dot{\avepos} - \omegagait \right] - \spring \avepos.
\end{eqnarray}
where $\omega_g=0$ since there is no geometric phase enclosed. Given that the moment of inertia is fixed as $I_0=m_sR^2$ in the null gait and $\alpha=2m_vR^2\theta_h$, the equation for the long-time envelope is, therefore

\begin{eqnarray}
I_0 \ddot{\avepos} &= - \frac{2 \cf m_s}{Tm_v|\ddot{\theta}_h|} \dot{\avepos} - \spring \avepos.
\end{eqnarray}

When $\tau_C$ is relatively small, the oscillation period of $\bar{\phi}$, $T_{\text{env}}$, is approximately $(2\pi/T_{\text{env}})^2=\tau_g/I_0$ where $\tau_g=m_sgR\sin{\theta_g}$ and $ I_0=m_sR^2$. This implies

\begin{eqnarray}
\theta_g\approx\sin{\theta_g}=\frac{R}{g}\left(\frac{2\pi}{T_{\text{env}}}\right)^2.
\end{eqnarray}

\begin{figure}[ht!]
  \centering
  \includegraphics[width=0.45\textwidth]{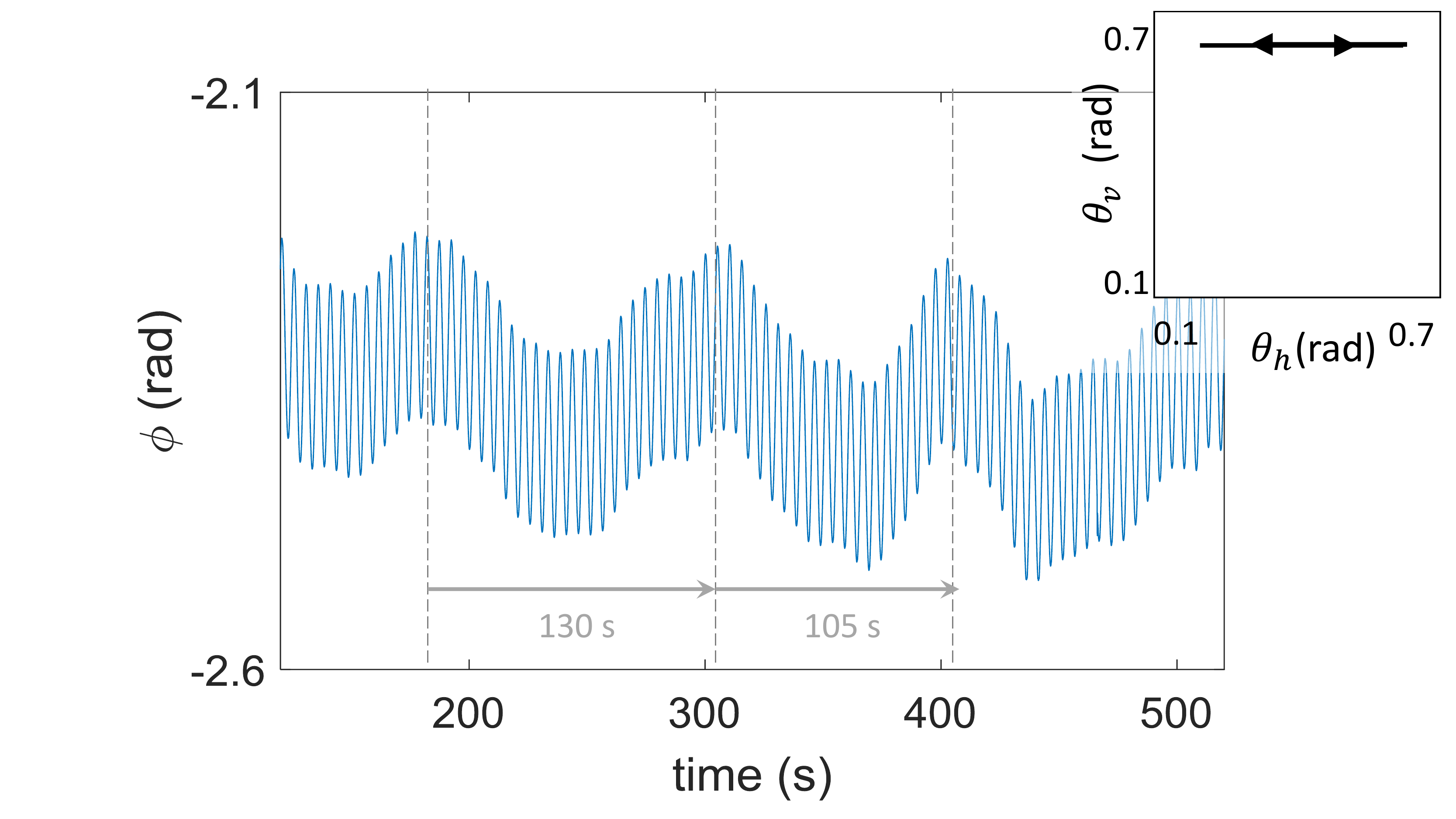}
  \caption{\textbf{Inferring the gravity residue.} The residual gravity angle is inferred from the long-time envelope of a null gait experiment with small friction ($\tau = 0.003 $ kg m$^2$ s$^{-2}$).}
\label{fig:residueInfer}
\end{figure}

With $R=0.46$ m, from an experiment with null gait and small friction $\tau = 0.003$ kg m$^2$ s$^{-2}$, we can see the period of the envelope $T_{\text{env}}$ is about $120$ s and thus inferring a residual gravity of $\theta_g = 1.2 \times 10^{-4}$ rad.

\section*{S6. Numerical integration}
To obtain the equation for numerical integration, we plug the angular momentum Eq.\ref{eq:angMom} into the equation of motion Eq.\ref{eq:EoM}, so
\begin{eqnarray}
-\tau_C\text{sgn}(\dot{\phi})-\tau_g\phi&=&\frac{dL}{dt}\\
&=&\frac{dI}{dt}\dot{\phi}+I\ddot{\phi}+\ddot{\alpha}\\
&=&\frac{\partial I}{\partial\theta_v}\dot{\theta}_v\dot{\phi}+I\ddot{\phi}+\ddot{\alpha}
\end{eqnarray}

Plugging in the $I(\theta_v(t))$ in Eq.\ref{eq:I} and the $\alpha$ in Eq.\ref{eq:alpha}, we finally arrive at

\begin{eqnarray}
\ddot{\phi}&=&I(\theta_v)^{-1}\nonumber\\
&&\left(2m_vR^2\sin{(2\theta_v)}~\dot{\theta}_v\dot{\phi}-2m_hR^2\ddot{\theta}_h-\tau_C\text{sgn}(\dot{\phi})-\tau_g\theta\right)\nonumber\\
\label{eq:simEq}
\end{eqnarray}

The numerical simulation integrates Eq.\ref{eq:simEq}. The initial position $\phi(0)=0$ such that the swimmer starts from the bottom of the slight potential well from the residual gravity. The initial angular velocity $\dot{\phi}$ is chosen that the initial angular momentum $L(0)=I(0)\dot{\phi}(0)+\dot{\alpha}(0)$ is zero.

The motor positions $\theta_v(t)$ and $\theta_h(t)$ use the commanded signal sent to the motor (see Fig.S1). The sign function in the Coulomb friction is smoothed by the arctan function with characteristic angular speed $0.01~\text{rad/s} \ll$ the typical speed of the swimmer to avoid the numerical singularity. There are two motors on the horizontal track and two motors on the vertical track. The mass of each motor is $0.116$ kg $=m_v=m_h$. The radius of the swimmer is $R=0.46$ m. The mass the track is $m_{\text{track}}=0.388$ kg.

For the convenience of implementing the same gait as the commanded shape change sent to the motors, which are discrete signals requiring interpolation in the differential equations to be integrated, we use a numerical scheme with fixed steps (forward Euler) so that the interpolation of the input signal is time-economic. The test of convergence with step size $h$ shows a global (i.e. position $\phi$) error of $O(h)$ (and therefore local error of $O(h^2)$) as expected for a first-order scheme. We use the step size $h=3.1\times 10^{-4}$ s such that the relative error is $1.8~\%$.

\begin{figure}[ht]
  \centering
  \includegraphics[width=0.45\textwidth]{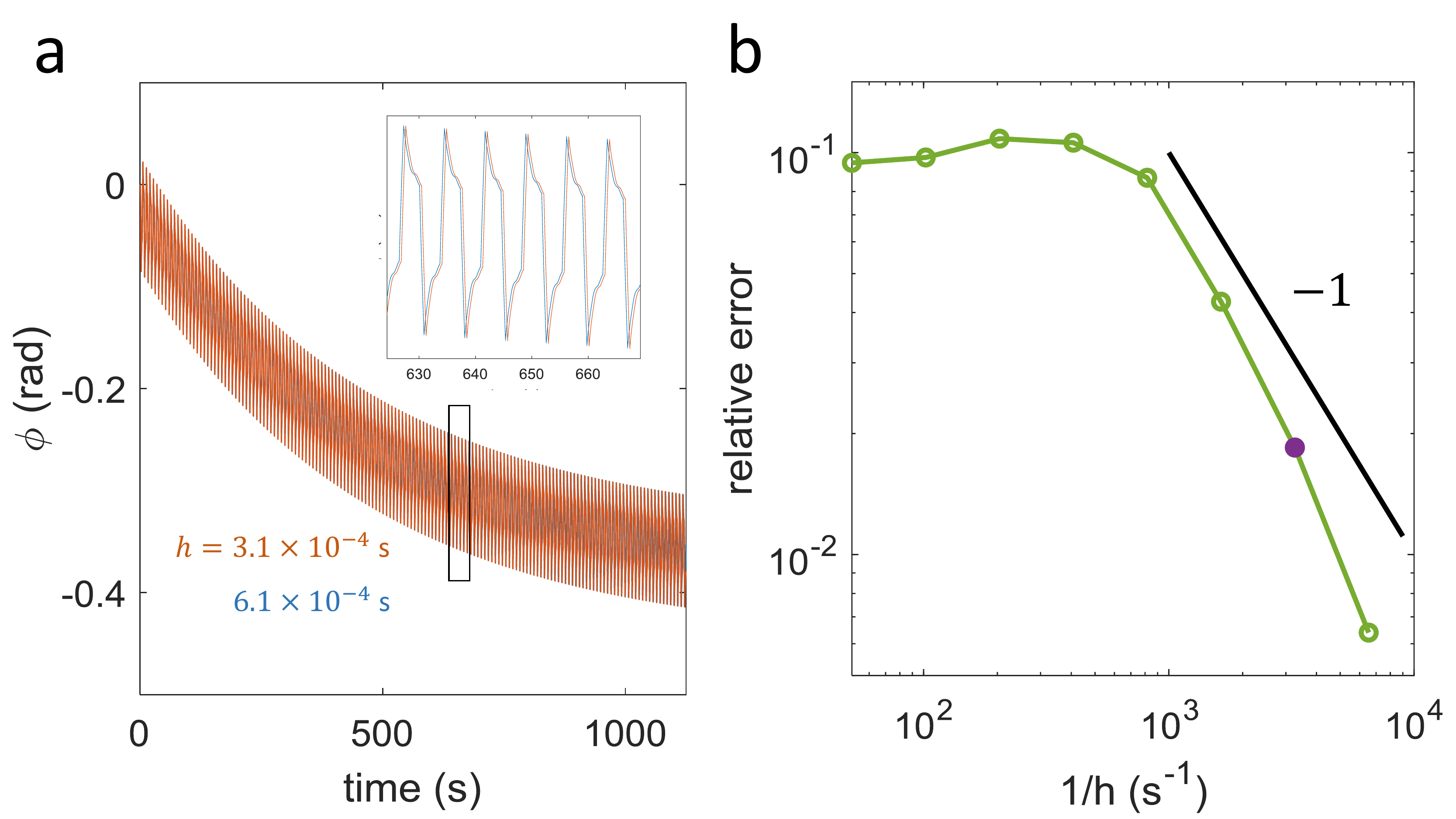}
  \caption{\textbf{Numerical convergence check.} (a) Numerical integration of a swimmer driven by a square swimming gait and subjected to a friction of $\tau_C=8.6\times 10^{-3}$ kg m$^2$s$^{-2}$ and residual gravity $\tau_g=4.0\times 10^{-4}$ kg m$^2$s$^{-2}$ with step sizes $h=3.1 \times 10^{-4}$ s and $6.1 \times 10^{-4}$ s. The inset shows a close-up at around $650$ s. (b) The relative error from numerical integration improves with the decrease of step size $h$. The improvement largely follows a trend of $O(h)$, which can be seen by a comparison with the black line that has a power of $-1$. Here the relative error is defined as the average of $|\phi_{\text{num}}-\phi_{\text{truth}}|/|\phi_{\text{num}}|$ where we use $h=7.7 \times 10^{-5}$ s to approximate $\phi_{\text{truth}}$. The step size we use in this study, $h=3.1\times 10^{-4}$, is shown in a solid purple dot.}
    \label{fig:negMomSim}
\end{figure}

\section*{S7. Plateau and negative momentum}

For small $L$, Eq.4 in the main text gives
\begin{eqnarray}
0&=&\frac{1}{\langle I\rangle}\left(-\tau_C\frac{4\bar{L}}{T|\ddot{\alpha}(0)|}-\tau_g\bar{\phi}\right)\\
\frac{L_{ss}}{\phi_{ss}}&=&-\frac{\tau_g T|\ddot{\alpha}(0)|}{4\tau_C}\\
&=&-\frac{g\sin{\theta_g} T|\ddot{\alpha}(0)|}{4Ra_C}\\
&=&-\frac{gR\sin{\theta_g} Tm_h |\ddot{\theta}_h(0)|}{2a_C}
\end{eqnarray}

To evaluate the angular momentum $L=I(t)\dot{\phi}+\dot{\alpha}$, we compute with all components ($\phi(t),I(t)=2m_v R^2 \cos^2{\theta_v(t)}+2m_h R^2+I_{\text{track}}R^2, \alpha(t)=2m_hR^2\theta_h(t)$) from experimentally recorded data. As the raw angular momentum is expected to have large fluctuation, we filter it with a Gaussian filter with width $\sigma = T$. The steady-state values of angular momentum and $\phi$, which are $L_{ss}$ and $\phi_{ss}$ use the average of the last $80\%$ in swimming stage since the incipient part is transient.

We also want to note that the characteristic high friction $\tau_{C0}=|\ddot{\alpha}|=2m_hR^2|\ddot{\theta}_h|$ where $|\ddot{\theta}_h|=0.12~\text{rad/s}^2$ (which can be read from Fig. \ref{fig:gaitExpt}).

In fact, if we consider a moment $t$ where $L(t)=\dot{\alpha}(t)$, then after a short time $\delta t$,
\begin{eqnarray}
\delta(I\dot{\phi})&=&\delta(L-\dot{\alpha})\\
&=&\dot{L}\delta t - \ddot{\alpha}\delta t.
\end{eqnarray}

For a large $\tau_C$ that we can neglect $\tau_g$ so that $\dot{L}=-\tau_C\text{sgn}(\dot{\phi})$, we have
\begin{eqnarray}
\delta(I\dot{\phi})=(\pm\tau_C-\ddot{\alpha})\delta t.
\end{eqnarray}

This indicates that when $\tau_C>|\ddot{\alpha}|$, the steady state $\phi$ cannot be further improved.

\begin{figure}[ht]
  \centering
  \includegraphics[width=0.45\textwidth]{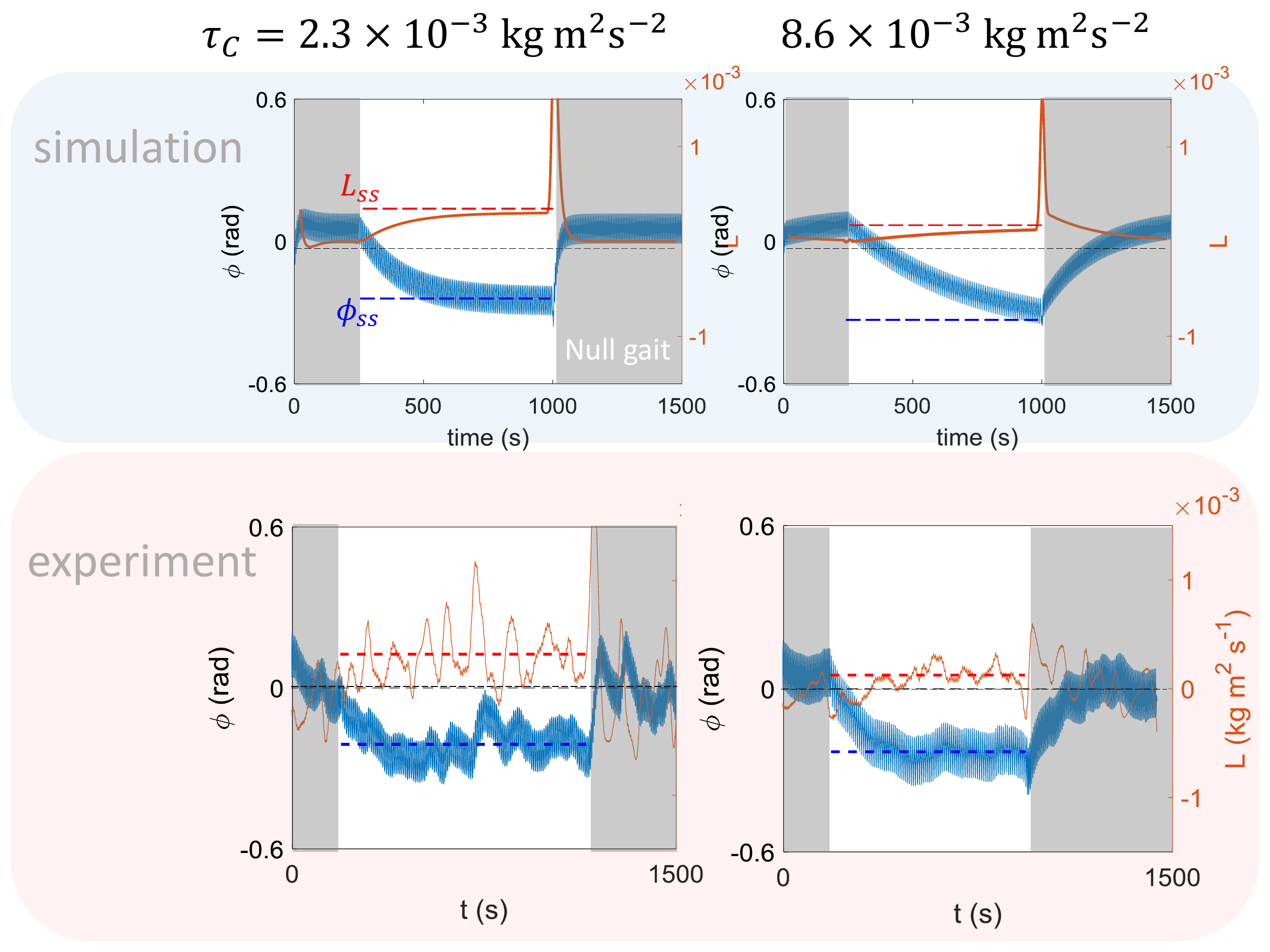}
  \caption{\textbf{Simulation of negative momentum.} Note that the $L_{ss}$ gets smaller with the torque of friction $\tau_C$.}
    \label{fig:negMomSim}
\end{figure}

\end{document}